\documentclass[letterpaper]{article} 
\usepackage{aaai24} 
\usepackage{times}  
\usepackage{helvet}  
\usepackage{courier}  
\usepackage[hyphens]{url}  
\usepackage{graphicx} 
\usepackage{natbib}  
\usepackage{caption} 
\usepackage{algorithm}
\usepackage{algorithmic}
\usepackage{xcolor}

\usepackage{newfloat}
\usepackage{listings}
\DeclareCaptionStyle{ruled}{labelfont=normalfont,labelsep=colon,strut=off} 
\floatstyle{ruled}
\newfloat{listing}{tb}{lst}{}
\floatname{listing}{Listing}
\nocopyright 

\title{Assessing the Impact of Case Correction Methods on the Fairness of COVID-19 Predictive Models}
\author {
    Daniel Smolyak,
    Saad Abrar,
    Naman Awasthi,
    Vanessa Frias-Martinez
}
\affiliations {
    University of Maryland, College Park\\
    dsmolyak@umd.edu, sabrar@umd.edu, nawasthi@umd.edu, vfrias@umd.edu
}

\usepackage{caption}
\usepackage{subcaption}
\usepackage{booktabs}
\usepackage{amsmath}
\usepackage{amssymb}

\begin{document}

\maketitle

\begin{abstract}
One of the central difficulties of addressing the COVID-19 pandemic has been accurately measuring and predicting the spread of infections. In particular, official COVID-19 case counts in the United States are under counts of actual caseloads due to the absence of universal testing policies. Researchers have proposed a variety of methods for recovering true caseloads, often through the estimation of statistical models on more reliable measures, such as death and hospitalization counts, positivity rates, and demographics. However, given the disproportionate impact of COVID-19 on marginalized racial, ethnic, and socioeconomic groups, it is important to consider potential unintended effects of case correction methods on these groups. Thus, we investigate two of these correction methods for their impact on a downstream COVID-19 case prediction task. 
For that purpose, we tailor an auditing approach and evaluation protocol to analyze the fairness of the COVID-19 prediction task by measuring the difference in model performance between majority-White counties and majority-minority counties. 
We find that one of the correction methods improves fairness, decreasing differences in performance between majority-White and majority-minority counties,
while the other method increases differences, introducing bias. While these results are mixed, it is evident that correction methods have the potential to exacerbate existing biases in COVID-19 case data and in downstream prediction tasks. Researchers planning to develop or use case correction methods must be careful to consider negative effects on marginalized groups.
\end{abstract}

\section{Introduction}

One of the many challenges of quelling the COVID-19 pandemic was accurately assessing the state of disease spread at any given moment. Particularly early in the pandemic, when testing was notoriously hard to access, official case counts were greatly under counting the actual spread of COVID-19. These undercounts led to confusion among everyday people, attempting to determine their level of risk for COVID-19 infection, and among public health officials, aiming either to allocate medical resources or to set social distancing policies. Even as testing capacity dramatically increased, many COVID-19 infections were not captured in official counts. Patients with asymptotic infections often did not get tested, and cases detected with at-home testing kits typically were not reported. 

To complicate matters further, racial bias has also been identified in COVID-19 data. 
For example, a lack of consistency in reporting race and ethnicity in COVID-19 statistics has generated a lot of missing or incorrect racial data, especially for minority racial and ethnic groups \cite{douglas2021variation}. 
Inadequate testing among minority racial and ethnic groups, such as Latino communities, has also impacted COVID-19 case statistics \cite{ama}; and 
official death data did not capture the excess deaths that were indirectly caused by COVID-19 – these deaths were 3 to 4 times more common among Black persons \cite{shiels2021racial}.

As under counting bias confounded with racial bias were being embedded into COVID-19 case and deaths counts, researchers have been creating
a plethora of predictive methods, from epidemiological models to machine learning approaches, to predict official COVID-19 case counts. 
In fact, the COVID-19 Forecast Hub is a good example of these efforts, with over $50$ independent teams uploading weekly predictions to a publicly available platform, and providing statistics to inform decision making \cite{Cramer2022-hub-dataset}.
However, despite a widespread understanding of the existence of under-counting and racial bias in COVID-19 data, researchers working on COVID-19 predictive methods often use publicly available - and potentially biased - case and deaths datasets to predict future cases and deaths \cite{arik2020interpretable, zhang2021seq2seq}.

To address under counting in COVID-19 datasets, researchers have developed \textit{case correction methods} that estimate ``true'' COVID-19 case counts. These estimates rely on statistical modeling from more reliable data, such as death and hospitalization counts and test positivity rates. While these metrics are themselves flawed, they are nevertheless more reliable than raw case counts.
In addition, due to the lack of disaggregate COVID-19 case data stratified by racial and ethnic groups \cite{kader2021participatory}, these correction methods are often applied to aggregate case counts, rather than to case counts per racial and ethnic group. As a result, it is unclear whether current case correction methods might introduce or amplify racial bias in case counts and in downstream prediction tasks.  

COVID-19 has disproportionately affected Black and Hispanic communities, as seen through increased rates of hospitalization and death \cite{mackey2021racial}. This is due to factors stemming from structural racism, such as decreased access to healthcare, increased housing density, and disproportionate representation in front-line occupations. Thus, it is important that case correction methods do not increase this burden through bias in corrected cases, and on downstream prediction tasks 
that use the corrected cases as training data. 

The main objective of this paper is to evaluate whether 
applying case correction methods to the COVID-19 datasets used to train predictive models, might impact the fairness of the predictions across racial and ethnic groups.
For that purpose, we propose a tailored auditing approach and evaluation protocol to analyze the impact of COVID-19 case correction methods on the fairness of COVID-19 county case prediction models that use corrected cases as training data, instead of \textit{official} counts from public sources such as NYT \cite{nytimes2021covid}. 
We focus on COVID-19 case prediction models at the county level, since these are closer to local realities and allow for more actionable decision making than state-level predictions. 
At its core, the audit trains and tests COVID-19 county case prediction models with \textit{official} (uncorrected) and corrected cases - obtained using two state of the art case correction methods - and evaluates changes in prediction fairness between models using official versus corrected cases from the two case correction approaches. 

We propose to measure the fairness of COVID-19 county case prediction models using the Accuracy Equality Ratio (AER) \cite{castelnovo2022clarification}, a fairness metric that measures the ratio between the prediction errors of protected and unprotected racial and ethnic groups (minority racial groups versus White). This choice is motivated by the need to guarantee that racial and ethnic minorities, who have been deeply impacted by the pandemic, are not further burdened by bias in case predictions. Social distancing policy decisions, such as restrictions on social gatherings or stay-at-home orders, are informed by projections of future caseloads. Inaccurate projections could lead to sub-optimal timing of the start and end of such policies. Prior research has shown that delays in lockdown starts led to higher cumulative caseloads \cite{huang2021impact}, while keeping lockdowns in place for longer than necessary can impose economic harms \cite{asahi2021effect}. Systematically worse prediction errors for marginalized communities could lead to mistiming both lockdown starts and ends, and generate disproportionate health and economic harms for those communities.

Due to systemic data collection failures during the pandemic \cite{kader2021participatory}, neither the official cases from public sources  nor the case correction methods we explore, provide the race-stratified case data at the county level that would allow us to predict cases by race, compute the errors, and measure the prediction fairness via the AER. 
Hence, we propose two approaches to assign county prediction errors to race or ethnic groups, and we evaluate whether the race assignment scheme impacts the fairness analysis. 
Finally, we focus on linear regression models for the COVID-19 case prediction task, given their simplicity and interpretability: epidemiological models are harder to tune due to their parametric nature and more complex deep learning models have black box architectures that are more difficult to interpret ~\cite{rudin2019stop}.

To summarize, the main contributions of this paper are:
\begin{itemize}
\item We tailor an auditing framework and evaluation protocol to analyze the impact of COVID-19 case correction methods on the fairness of county case prediction models, with a focus on linear regression approaches. 

\item We replicate and adapt state of the art COVID-19 case correction methods to the U.S. context, with a focus on county case statistics. The two methods follow distinct approaches: one method uses past COVID-19 deaths to correct case values while the other uses both cases and deaths to output corrected trends in caseloads.

\item We use the auditing approach to empirically evaluate the effect of the two case correction methods on the fairness of linear COVID-19 county case prediction models. Our results show that the correction method based only on death data exacerbates bias (decreases fairness) in the prediction task after cases are corrected, while the case correction method based on case and death data appears to at least maintain, and sometimes improve, the fairness of the predictions.

\end{itemize}

\section{Related Work}

\subsection{Fairness Correction in Regression Settings}
To correct for bias and unfair performance in predictive models, researchers have used pre-processing and in-processing correction approaches. Pre-processing approaches focus on removing bias from datasets prior to input into predictive models \cite{GlOVEPreFair, OptoPre, zhang2016causal}.
On the other hand, in-processing approaches to improve the fairness of predictive models are generally applied during the training process, penalizing predictions that are dependent on protected attributes. For example, in-processing approaches for linear models often add fairness constraints \cite{agarwal2019fair} or fairness regularization to the objective \cite{berk2017convex}, and in-processing approaches for deep learning architectures usually add a bias correction term in the loss function \cite{inproc_Socmedia, Das_Dooley_Inproc, Yan_Seto_Apostoloff_inproc}. In this paper, we focus on pre-processing approaches that correct COVID-19 case datasets prior to the prediction task.

\subsection{Pre-processing Case Correction Methods}
\label{sec:case-correction-methods}

Researchers have developed a variety of methods to estimate true COVID-19 case counts. \cite{jagodnik2020correcting} used a benchmark country with high testing capacity (in their case South Korea) to estimate the case fatality rate (CFR). They then adjusted the CFR for other countries based on relative population age, and used deaths as ground truth to estimate case numbers. Similarly, \cite{albani2021covid} used estimates of rates of hospitalization and death after infection in a ``stable'' period of the pandemic when testing was more widely available to back-estimate early caseloads. Furthermore, they analyzed the impact of using corrected versus uncorrected cases to inform a vaccination campaign on the effectiveness of the campaign. \cite{basu2020estimating} uses an exponential decay model to estimate the infection fatality rate at the U.S. county level, and researchers in \cite{schneble2021statistical} estimated a dynamic mixed regression model to find trends in the infection fatality rates, and thus, infection rates. In \cite{wu2020substantial}, the authors created a simulation framework for COVID-19 caseloads by U.S. state, using probabilistic bias analysis and data on test positivity and test accuracy. \cite{li2020substantial} also simulated COVID-19 spread, using epidemiological models and mobility data to infer both epidemiological parameters (such as reproduction rate) and numbers of infections. \cite{nightingale2022local} estimated infection detection rates in England in a time period before testing expansion by using the infection detection rates post expansion in a Bayesian mixed effects models. Similarly to \cite{jagodnik2020correcting}, they also incorporated age and a deprivation index into their model to account for differences in vulnerability to COVID-19 across the regions of interest. \cite{nicholson2022improving} used 
Bayesian models and Monte Carlo sampling to estimate prevalence from a randomized COVID-19 surveillance study, and they made use of a causal framework to conduct debiasing of prevalence data.

Another common method of finding infection rates is ``seroprevalance'' studies, where blood samples are directly analyzed for evidence of COVID-19 infection. One seroprevalence study found that infection rates were 3 to 10 times higher than reported in various periods of 2020 \cite{angulo2021estimation}. However, such studies are slow to adapt to changing infection dynamics. In our study, we choose to focus on \cite{jagodnik2020correcting} and \cite{schneble2021statistical} because a) their code is publicly available and easily adaptable to the U.S. county context (\textit{e.g.,} do not require further data sources) and b) they are representative of the broader pool of approaches. \cite{jagodnik2020correcting} represents both the relatively simpler correction methods developed in early phases of the pandemic and the use of population demographics (only age) for correction, while \cite{schneble2021statistical} represents the more complex statistical (and often Bayesian) models developed in later phases, and provides coverage of a longer study period.

\subsection{Fairness Metrics in Regression Settings}
As with fairness metrics in classification settings, fairness metrics in regression settings aim to measure the disparate impacts of regression models on protected attributes or groups. These metrics typically focus either on the independence of the protected attribute and the predictions \cite{agarwal2019fair, steinberg2020fairness, silvia2020general, calders2013controlling, du2022fair} or the protected attribute and the model performance \cite{gursoy2022error, zink2020fair}. 
For this paper, we make use of the latter approach, and focus on the relationship between the prediction errors for protected and unprotected groups via the Accuracy Equality Ratio (AER) \cite{castelnovo2022clarification}, 
a metric that measures the ratio between the prediction errors for protected and unprotected racial and ethnic groups.

\section{Data}

\begin{figure}
    \centering
    \includegraphics[scale=0.4]{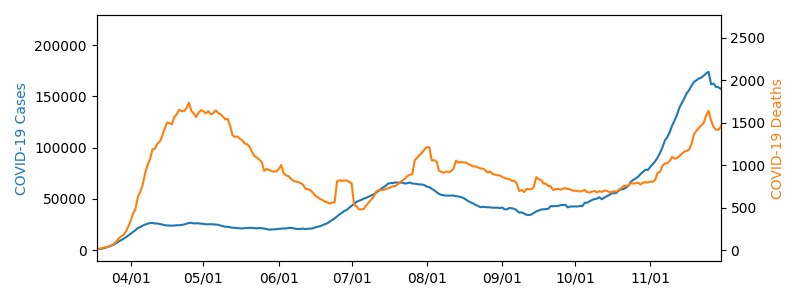}
    \caption{Daily cases and deaths in the United States in 2020, 7-day rolling average.}
    \label{fig:us-covid-2020}
\end{figure}

We acquire COVID-19 case and death data by U.S. county from the New York Times data repository\footnote{\url{https://github.com/nytimes/COVID-19-data}}. To the best of our knowledge, this dataset has not been corrected for case under counts. Average daily cases and deaths are reported as 7-day rolling averages, to smooth the impact of irregular reporting practices whereby cases were reported or submitted every few days. 
For our analysis, we use COVID-19 case and death data from March 18th, 2020 to November 30th 2020, as shown in Figure \ref{fig:us-covid-2020}. We focus on this 258-day period to analyze the height of the pandemic prior to the availability of COVID-19 vaccines.

\section{Case Correction Methods}
\label{correction_methods}
Our objective is to evaluate whether case correction methods applied to COVID-19 datasets prior to the training of prediction models have any impact on the fairness of county predictions in the U.S.  
Next, we describe the two case correction methods we focus on, and how we adapt them to our setting.

\subsection{Method 1: Dynamics in Infection Numbers}
The first case correction method we examine is  \cite{schneble2021statistical}, which aims to estimate relative changes in the infection numbers over time. They assume the case fatality ratio (CFR), $a$ in their notation, is constant, but they also seek to estimate the relative case detection ratio, $c_t$. To do so, \cite{schneble2021statistical} estimates the following dynamic mixed regression model with smooth random effects variable, $V_t$, such that $V_t=\text{log}(I_t)$, where $I_t$ is the number of infections at time $t$

\begin{equation}
\mathbb{E}(Y_t | V_t, x) = exp(V_t + \beta_0 + x\beta_t) \text{ and } V_t \sim N(\mu_t, \sigma^2)
\end{equation}

where $x$ is either 0 or 1, and $Y_t$ is set to the reported number of fatal cases, $F_t$, or the number of reported recovered cases, $R_t$, when $x=0$ and $x=1$, respectively. As a result, $V_t$ provides the relative number of log-scale infections over time, starting at 0 on the first day. 
Although $V_t$ does not provide absolute infection numbers, due to identifiability issues in the regression, 
we regardless use $V_t$ directly as our ``corrected'' cases metric because the authors do recover the absolute infection numbers in their paper \cite{schneble2021statistical}. 

\subsection{Method 2: CFR Benchmark}
The second method we consider is from \cite{jagodnik2020correcting}, which takes a relatively straightforward approach of finding adjusted cases from reported death numbers. To do so, they first find a Case Fatality Ratio (CFR) from a benchmark country, $B$, where testing is more widely available – \cite{jagodnik2020correcting} selects South Korea in their analysis. They then compute a Vulnerability Factor to account for the different age distributions of the target country, $T$, and the benchmark country, and the varying CFR by age group, $r_i$.

\begin{equation}
V_{TB} = \frac{\sum_{i=0}^{N} f_{T_i} r_i}{\sum_{i=0}^{N} f_{B_i} r_i}
\end{equation}

The proportion of the population of country $T$ ($B$) in age group $i$ is $f_{T_i}$ ($f_{B_i}$). A vulnerability factor greater than 1 indicates an older population in the target country, with greater vulnerability to fatality from COVID-19. We now find the adjusted cumulative cases, $AC$, for day $t$ as the ratio between reported deaths, $D$, offset by a delay between infection and death, $d$, and the CFR and vulnerability factor.

\begin{equation}
AC(t) = \frac{D(t+d)}{V_{TB} \times CFR_{B}}
\end{equation}

There are a few differences between \cite{jagodnik2020correcting} and our adaptation. First, we use counties as our target regions as opposed to countries - we are still able to calculate vulnerability factors as county age demographics are available from the U.S. Census. Next, we use a different time period of evaluation - \cite{jagodnik2020correcting} examines the start of the pandemic up until the end of March, around the time of the release of their method. Because few U.S. counties have substantial case or death data in March, we expand our analysis to include April and May of 2020. Finally, we use reported future deaths in our adjusted case calculation while \cite{jagodnik2020correcting} uses predicted future deaths. This helps avoid the complication of additionally evaluating their method for predicting future deaths.

\section{Audit Structure: Impact of Case Correction Methods on Prediction Fairness}
In this section, we present our methods for auditing the impact of COVID-19 case correction methods 
on the fairness of county case predictions models, describing the main steps. 
We will cover how this structure is used for our empirical evaluation in the next section. 

\subsection{Step one: Compute Corrected COVID-19 cases}
We first apply the two case correction methods described in the previous section: dynamics in infection numbers (method 1) and CFR Benchmark (method 2) to publicly available, \textit{official} NYT data for COVID-19 county cases during the period of interest. Running the two correction methods will output two corrected distributions, one per case correction approach.
Details about how these methods are applied will be described in the next section: evaluation protocol.

\subsection{Step two: Train Quantile Regression Models}
\label{sec:quantile-regression}

Quantile regression is a type of regression analysis that estimates the conditional median (or other quantiles) of the response variable (COVID-19 cases in our context), instead of the conditional mean used in least squares estimates. 
We use quantile regression as the COVID-19 county case prediction model since, unlike least square estimates, quantile regression models are more robust against outliers. 

The independent and dependent variables in our quantile regression model are all daily county caseloads (official or corrected). Given a prediction at day $d$ and a time series list of cases $C$, the predictors for the model are the cases numbers for the 3 previous days, $C_{d-1}, C_{d-2}$, and $C_{d-3}$. We use 3 days to balance the sensitivity to outliers from using fewer days and the decreasing information gained from additional previous days. The day we aim to predict, given a lookahead $l$ is $C_{d+l}$.

\begin{equation}
    C_{d+l} \sim \beta_0 + \beta_1 C_{d-1} + \beta_2 C_{d-2} + \beta_3C_{d-3}
\end{equation}

We use the Python library \textit{scikit-learn} to estimate this equation as a quantile regression \cite{scikit-learn}. 
After dividing the time period under study into a ``train'' and ``test'' period, we fit a quantile regression on the training period data for each county, with quantiles of 0.05, 0.5, and 0.95. This is repeated on both the original cases, as our baseline, and the corrected cases obtained in step one. Additionally, to explore the effect of how far in the future we aim to predict cases, we repeat the model training over multiple ``lookaheads'' (e.g., 1 day ahead, 1 week ahead, 2 weeks ahead, etc). 

\subsection{Step Three: Compute Prediction Error}
In the third step, we assess the performance of the quantile regression model on each test dataset, taking the average pinball loss (PBL) between the predicted cases and the test cases over the three quantiles. To account for differences in county population (e.g., an error of 100 cases has vastly different implications in a county of ten thousand residents versus a county of ten million residents), we divide each county's PBL by its population. However, we do not population-adjust the PBLs for Method 1's corrected data, as Method 1 produces a trend and not case counts.

Hence, the pinball loss $L_{\tau}(y, f_\tau)$ is computed as: 

\begin{equation}
L_{\tau}(y, f_\tau) = 
\begin{cases} 
\tau (y - f_\tau) & \text{if } y \geq f_\tau, \\
(\tau - 1) (f_\tau - y) & \text{if } y < f_\tau
\end{cases}
\end{equation}

where $L_{\tau}(y, f_\tau)$ denotes a county's pinball loss for a given quantile $\tau$, $y$ is the observed value, and $f_\tau$ is the forecasted value at quantile $\tau$. For our analysis, we use the average county pinball loss (PBL), computed across the set of $3$ quantiles, $T$, and normalized by the county population. 
$$
\overline{PBL} = \frac{1}{|T|} \sum_{\tau \in T} L_{\tau} (y, f_\tau)
$$

At the end of this step, we have two PBLs for each county and for each lookahead: one for the corrected case setting and one for the uncorrected case setting. With these PBLs we can construct our fairness analysis.

\subsection{Step Four: Assignment of County Racial/Ethnic Label}
\label{sec:race_assign}

To carry out the fairness analysis, we focus on the fairness metric Accuracy Equality Ratio (AER) that measures the ratio between the prediction errors of protected and unprotected racial and ethnic groups. This choice is motivated by the need to guarantee that racial and ethnic minorities, who have been deeply impacted by the pandemic, are not further burdened by bias in case predictions, which could in turn affect how social distancing policy decisions are made. 

Hence, for the fairness analysis, we first need to associate the county prediction errors (PBL) with the county's racial and ethnic groups since our fairness metric is focused on race-based prediction errors.
However, that would require access to race-stratified COVID-19 case data at the county level, unfortunately not available due to systemic data collection failures during the pandemic~\cite{kader2021participatory}. 
Thus, we propose to assign each county a single racial or ethnic group label with respect to their composing racial and ethnic groups. To explore the effects of different assignment schemes, we investigate both plurality and majority assignments.

For the majority assignment, we select the racial or ethnic group that composes $\ge50$\% of the county population – if no group holds a majority then we assign a label of ``Non-White''. For example, if a county is 40\% White, 30\% Black and 30\% Hispanic, we consider its majority label ''Non-White'', as no group exceeds 50\%. For the plurality assignment, we select the racial or ethnic group in the county with the highest proportion regardless of whether that proportion is $\ge50$\% (in the prior example, this county's plurality assignment would be ``White''). 
We use population race and ethnicity data from the 2019 U.S. Census. We define the following group abbreviations from Census definitions: Asian is  ``Asian alone or in combination population''; Black is ``Black or African American alone or in combination population''; Hispanic is ``Hispanic population'' (all races); and White is ``Not Hispanic, White alone population''.

\subsection{Step Five: Compute Fairness Metrics}
Once a label is assigned to each county according to its majority or plurality racial and ethnic composition, we find the average PBL across the counties for each race/ethnicity label. Then, we calculate the fairness metric, Accuracy Equality Ratio (AER) for a given protected group as the ratio between the PBL for the majority/plurality counties for that protected race/ethnicity and the PBL for the majority/plurality White counties (unprotected group). 

Below, the AER for protected group $r$ (Asian, Black, Hispanic or Non-White) is $AER_r$;  the set of all counties for a given majority/plurality protected racial group $r$ is $C_r$; the set of all majority/plurality White counties is $C_w$, and the PBL for each county is $PBL_c$.

\begin{equation}
AER_r = \frac{\frac{1}{|C_r|}\displaystyle\sum_{c \in C_r} PBL_c}{\frac{1}{|C_w|} \displaystyle\sum_{c \in C_w} PBL_c}
\end{equation}

AERs above one are associated with higher errors (lower accuracy) for Asian, Black, Hispanic or non-White counties compared to White; and AERs below
one correspond to higher errors for majority/plurality White counties when compared to other protected racial groups.
The farther away the AER is from one for a given prediction model, the less fair the model is, since the difference (ratio) between the PBLs for protected and unprotected racial groups is larger.
An AER over 1 is particularly of concern given historic and structural racial discrimination, and the outsize effect of the COVID-19 pandemic on Black and Hispanic communities \cite{mackey2021racial}. AERs close to one reveal similar distribution of prediction errors for protected and unprotected groups, and point to fair prediction models. 

\section{Evaluation Protocol}
\label{evaluation}

We employ the audit structure described in the previous section to evaluate the impact of the two COVID-19 case correction methods on the fairness of county case predictions models in the U.S. 

For that purpose, we first apply the two case correction methods, infection dynamics correction and CFR benchmark, on the \textit{official} NYT case data, to obtain two datasets containing the corrected COVID-19 cases using each of the correction approaches. 
While the infection dynamics correction does not make any assumption about the COVID-19 evolution, the CFR method was created to be applied to exponential peak periods. Hence, we next describe how these model restrictions inform how we apply the methods to obtain the corrected datasets.

\begin{figure*}
     \centering
     \begin{subfigure}[b]{0.48\textwidth}
         \centering
         \includegraphics[width=\textwidth]{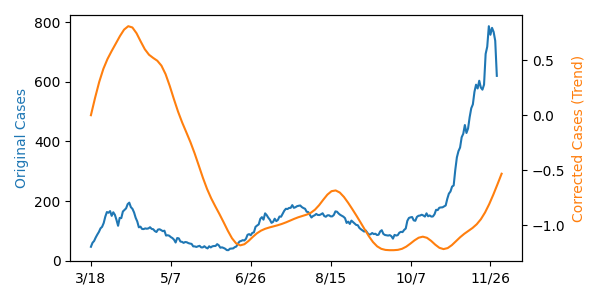}
         \caption{Method 1: Original cases for Kings County, Washington versus the ``corrected'' trend.}
         \label{fig:dynamic-example}
     \end{subfigure}
     \hfill
     \begin{subfigure}[b]{0.48\textwidth}
         \centering
         \includegraphics[width=\textwidth]{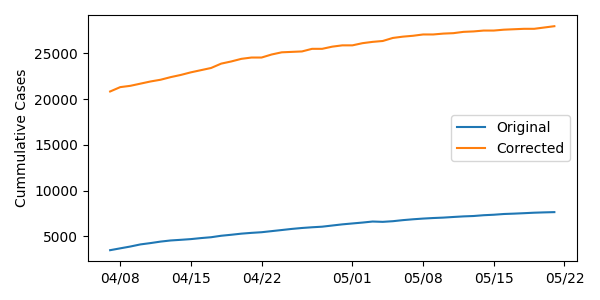}
         \caption{Method 2: Original versus corrected cumulative cases for Kings County, Washington.}
         \label{fig:cfr-example}
     \end{subfigure}
     \caption{County examples of original versus corrected cases.}
\end{figure*}

\begin{table*}[]
    \centering
    \begin{tabular}{ccccc|cccc}
        & \multicolumn{4}{c@{}}{\textit{Majority}} & \multicolumn{4}{c@{}}{\textit{Plurality}} \\ \cmidrule(l){2-9} 
         & Black & Hispanic &  Non-White & White & Asian & Black & Hispanic & White \\ \hline
            \textit{Method 1} & 85 & 71 & 125 & 1650 & 5 & 109 & 88 & 1729 \\
        \textit{Method 2 (4/7 - 5/22)} & 11 & 8 & 43 & 164 & 2 & 20 & 14 & 190 \\
        \textit{Method 2 (4/24 - 6/15)} & 26 & 13 & 63 & 410 & 3 & 39 & 21 & 449 \\
    \end{tabular}
    \caption{Frequency of counties per group assignment for each method, divided by study period for Method 2 (all dates in 2020).}
    \label{tab:demo_assignment}
\end{table*}

\textbf{Method 1: Dynamics in Infection Numbers.}
With the R code available from \cite{schneble2021statistical}, we estimated the relative number of infections for 2,869 U.S. counties from the official NYT reported cases and deaths for each of those counties in the 258 day period from March 18th, 2020 to November 30th 2020.
Figure \ref{fig:dynamic-example} show an example of official and corrected cases. We then excluded counties where either fewer than 5 deaths occurred during the chosen time period or where the estimated relative infections maintained a constant trend, as measured by the difference between the maximum and minimum slope between any two consecutive dates. This eliminated 936 counties, leaving 1,933 counties for our analysis. This ensured a sufficiently ``challenging'' prediction task for the quantile regression in order to assess fairness. Table \ref{tab:demo_assignment} provides the frequency of demographic assignments for these counties. We use a 70/30 split, taking the first 180 days as the training period and the remaining 78 days as the testing period.

\textbf{Method 2: CFR Benchmark.}
Because this method is designed for a period of exponential rise and decay in caseloads, we focus our testing period on the first phase of the pandemic in the United States. While \cite{jagodnik2020correcting} used the time period of January 22nd through March 28th (their date of publication), the vast majority of U.S. counties had no recorded deaths even through the end of March. Thus, we first focus on the 45-day time period from April 7th to May 22nd, taking the first 15 days as our train period and the remaining days as our test period. To ensure that we use the correction method properly, we only include counties with at least 5 cumulative deaths by April 7th, leaving 226 counties. Because of this low total, we sought to expand the number of counties included through the additional investigation of a second time period. Thus, we also examine the 44 day time period from April 24th to June 15th. After using the same cumulative death threshold for April 24th, 515 counties are included in the analysis. The distribution of counties among each demographic assignment is shown in Table \ref{tab:demo_assignment} for each study period, and an example of original versus corrected cases is shown in Figure \ref{fig:cfr-example}.

We then use the corrected training and testing datasets output by each method to train the quantile regression models. 
For Method 1, we train two quantile regression models, one with the official, uncorrected case data, and one with the corrected cases, to predict COVID-19 county cases for lookahead 7 (one week). We repeat this model training approach for the other three lookaheads (2, 3 and 4 weeks), training a total of eight models across four lookaheads and two case datasets per county.
For Method 2, we follow the same approach. However, given the limited training data, we only consider lookaheads 1 (one day), 7 (1 week) and 14 (2 weeks), computing a total of six models per county.
We refer to the models trained with uncorrected data as baseline models, since we will be comparing their performance against that of models trained with corrected cases. 
Once the models are trained, the PBL - stratified by racial and ethnic groups - is computed for each trained predictive model and lookahead,
with race and ethnicity labels assigned using either majority or plurality schemes.
Finally, the PBLs are used to compute the corresponding fairness metrics (AER) for each regression model.

\begin{figure*}[ht!]
     \hspace{-0.5cm}
     \begin{subfigure}{0.74\textwidth}
         \centering
        \includegraphics[scale=0.5]{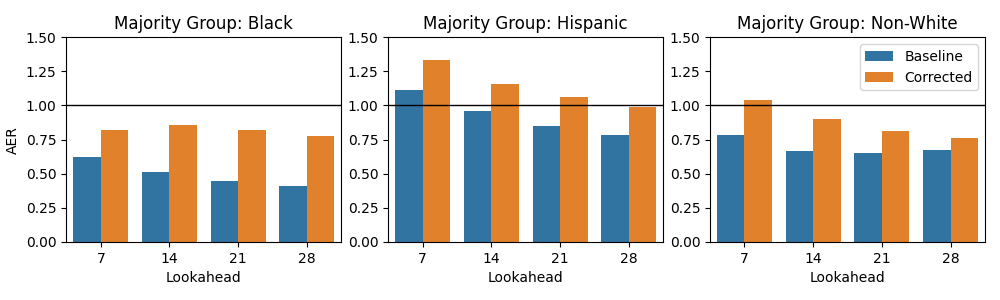}
        \caption{}
        \label{fig:dynamic-AER-majority}
     \end{subfigure}
     \hspace{-0.5cm}
     \begin{subfigure}{0.24\textwidth}
        \centering
         \includegraphics[scale=0.5]{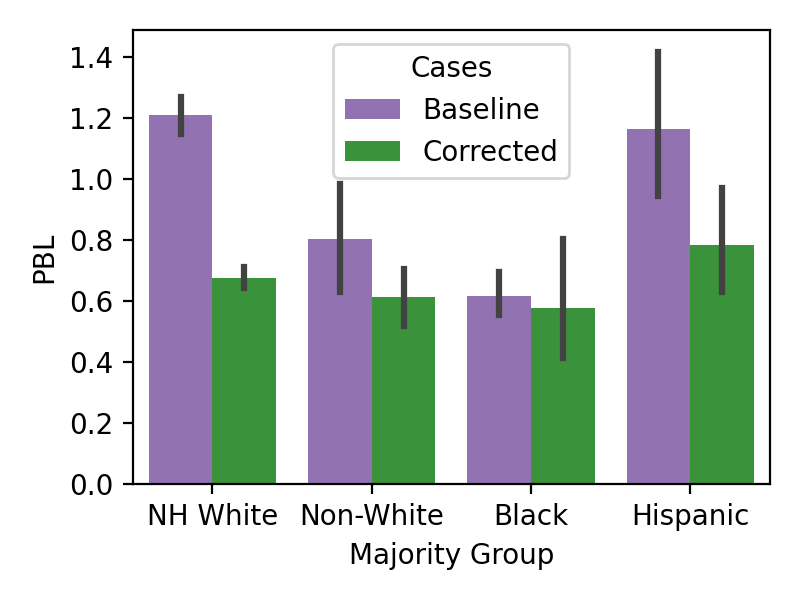}
         \caption{}
         \label{fig:dynamic-PBL-l14-maj}
     \end{subfigure}
     \caption{Method 1: a) AER values for Majority Black, Hispanic, and Non-White counties compared to majority-White counties for both uncorrected and corrected cases. b) The average and 95\% confidence interval of PBL values for each majority group for lookahead 14. Baseline PBLs are all multiplied by the median county population (25,726) and Corrected PBLs are scaled by 10 to improve readability.}
\end{figure*}

\begin{figure*}[ht!]
     \hspace{-0.5cm}
     \begin{subfigure}{0.74\textwidth}
         \centering
        \includegraphics[scale=0.5]{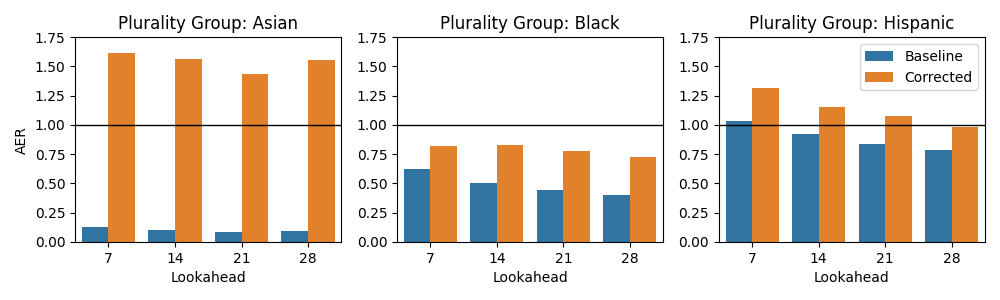}
        \caption{}
        \label{fig:dynamic-AER-plurality}
     \end{subfigure}
     \hspace{-0.5cm}
     \begin{subfigure}{0.24\textwidth}
        \centering
         \includegraphics[scale=0.5]{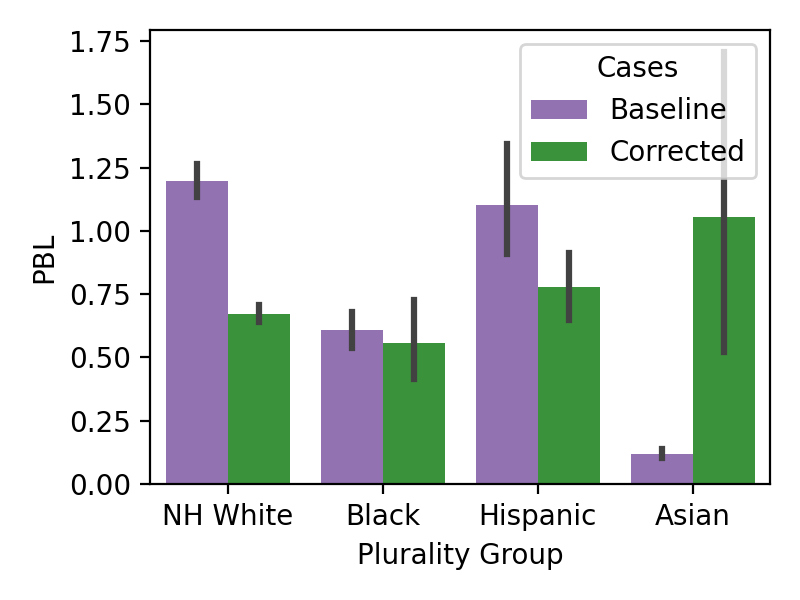}
         \caption{}
         \label{fig:dynamic-PBL-l14-plu}
     \end{subfigure}
     \caption{Method 1: a) AER values for Plurality Asian, Black, and Hispanic counties compared to plurality-White counties for both uncorrected and corrected cases. b) The average and 95\% confidence interval of PBL values for each plurality group for lookahead 14. Baseline PBLs are all multiplied by the median county population (25,726) and Corrected PBLs are scaled by 10 to improve readability.}
\end{figure*}
 
\textbf{Fairness Analysis.} 
To assess the impact of the correction methods on the fairness of the predictions across racial and ethnic groups, lookaheads, and race assignment scheme we propose to carry out comparative and statistical analyses between the AERs of the baseline model and the AERs of the model trained with corrected data (either from the infection dynamics method or the CFR benchmark method). 

In the \textbf{comparative analysis}, we explore whether AERs are getting closer to one from baseline to corrected models, pointing to an improvement in fairness \textit{i.e.,} prediction errors are more similar across protected and unprotected racial and ethnic groups. Additionally, we will discuss PBLs and changes in AER from smaller than one to larger than one and vice versa, when comparing baseline and corrected model, pointing to how prediction errors might be larger for protected racial and ethnic groups (AER$>$1) or for the unprotected group (AER$<$1), although the fairness, measured as absolute distance to one, might be similar. This is important to identify cases where, although our fairness metric might not change, the burden of the prediction errors shifts from protected to unprotected groups, and vice versa.

In the \textbf{statistical analysis}, we evaluate whether the fairness changes identified in the comparative analysis are statistically significant.  
For that purpose, we compare the PBLs of the White counties with the PBLs of each of the other racial or ethnic groups, utilizing the Mann-Whitney U test, and analyze whether the significance level of the test changes from the baseline to the corrected models. 
Going from significant tests for the baseline to non-significant tests for the corrected, will reveal an increase in fairness (similar PBL for protected counties and White counties), while the opposite will point to a fairness decrease.  
This test is conducted separately for each lookahead. Mann-Whitney is selected due to PBLs not being normally distributed.

\section{Results}

\begin{figure*}[ht!]
     \hspace{-0.5cm}
     \begin{subfigure}{0.74\textwidth}
         \centering
        \includegraphics[scale=0.5]{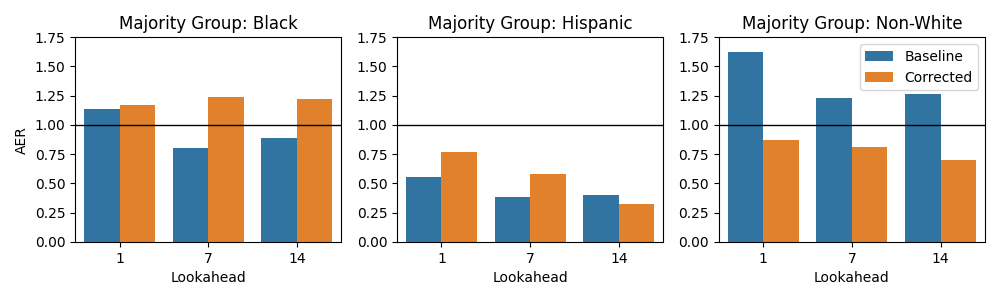}
        \caption{}
        \label{fig:cfr-AER-majority-225}
     \end{subfigure}
     \hspace{-0.5cm}
     \begin{subfigure}{0.24\textwidth}
        \centering
         \includegraphics[scale=0.5]{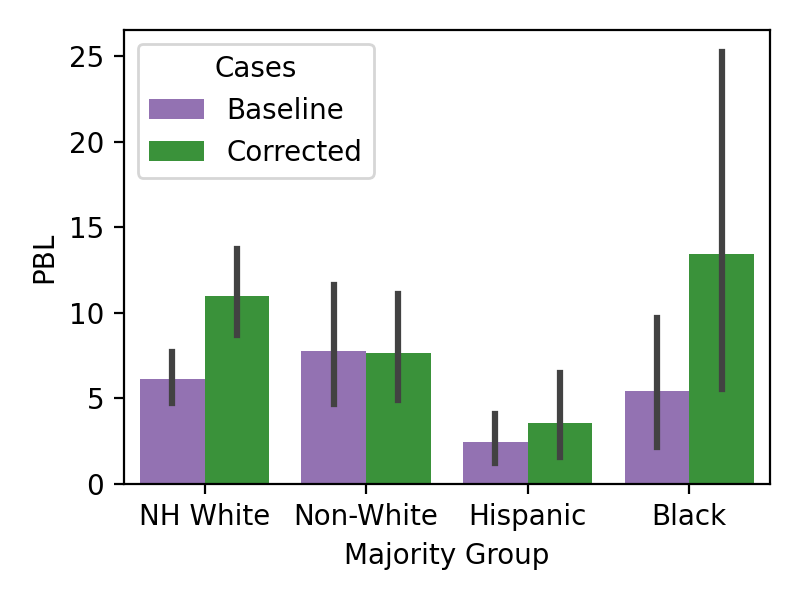}
         \caption{}
         \label{fig:cfr-225-PBL-l14-maj}
     \end{subfigure}
     \caption{Method 2, study period 4/7/20-5/22/20: a) AER values for Majority Black, Hispanic and Non-White counties compared to majority-White counties for both uncorrected and corrected cases. b) The average and 95\% confidence interval of PBL values for each majority group for lookahead 14. PBLs are all multiplied by the median county population (25,726).}
\end{figure*}

\begin{figure*}[ht!]
     \hspace{-0.5cm}
     \begin{subfigure}{0.74\textwidth}
         \centering
        \includegraphics[scale=0.5]{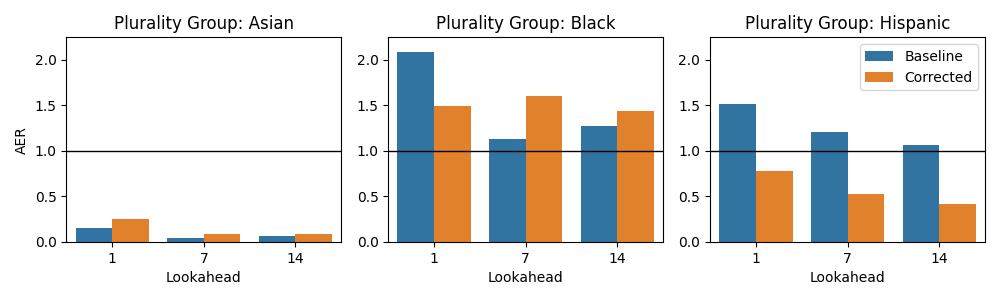}
        \caption{}
        \label{fig:cfr-AER-plurality-225}
     \end{subfigure}
     \hspace{-0.5cm}
     \begin{subfigure}{0.24\textwidth}
        \centering
         \includegraphics[scale=0.5]{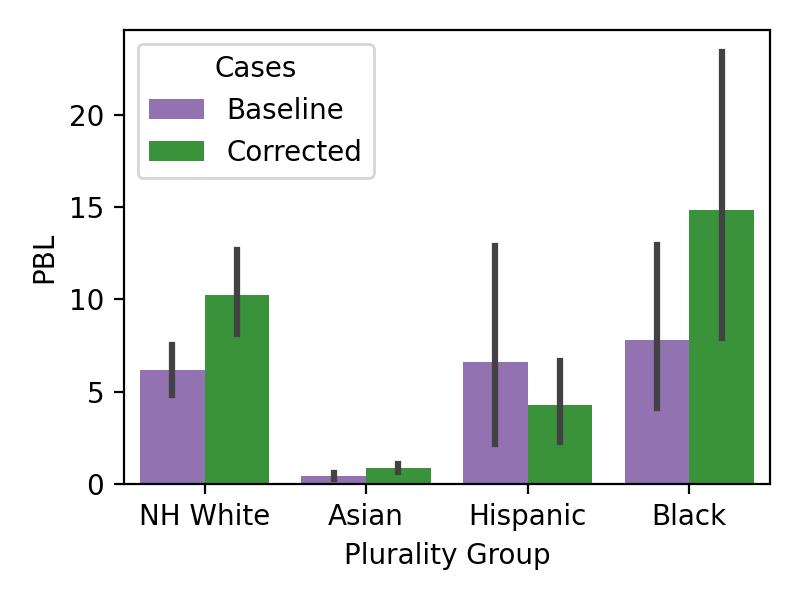}
         \caption{}
         \label{fig:cfr-225-PBL-l14-plu}
     \end{subfigure}
     \caption{Method 2, study period 4/7/20-5/22/20: a) AER values for Plurality Asian, Black, and Hispanic counties compared to plurality-White counties for both uncorrected and corrected cases. b) The average and 95\% confidence interval of PBL values for each majority group for lookahead 14. PBLs are all multiplied by the median county population (25,726).}
\end{figure*}

\subsection{Method 1: Impact of Dynamics in Infection Numbers on Quantile Regression Fairness}

The resulting AERs for the majority-assignment scheme are shown in Figure \ref{fig:dynamic-AER-majority}. For the baseline, we see that for majority-Black and majority-Non-white counties the AER is consistently below 1, indicating lower PBLs for those protected groups when compared to majority-White counties. 
Applying the Dynamic in Infection Numbers case correction for these two groups increases fairness across all lookaheads \textit{i.e.,} the corrected AERs are all closer to 1, and many of these improvements in fairness are statistically significant (as seen in Table \ref{tab:mwu-dynamic} in the Appendix).
For several of the lookaheads, baseline PBLs for Black- and Non-White-majority counties are statistically significantly lower than PBLs for White-majority counties, while corrected PBLs are not longer significantly different between racial/ethnic groups, pointing to higher fairness. 
Majority-Hispanic counties, on the other hand, have baseline PBLs starting above 1 for lookahead 7 and decreasing below 1 as the lookahead increases to 14 and above. 
For this protected group, the fairness decreases for lookaheads 7 and 14 while increasing for lookaheads 21 and 28. \textbf{The picture here is relatively clear that the correction method rarely adds bias to the modeling process, and very interestingly, it can even improve the prediction fairness across racial groups as measured by AER.}

The AERs calculated by plurality assignment are shown in Figure \ref{fig:dynamic-AER-plurality}. 
Plurality-Black county AER is closer to 0.5 for the baseline and moves towards an AER of 1 (without going over) when cases are corrected, pointing to an improvement in fairness, with PBL error distributions being closer between the plurality-Black and plurality-White counties.  Plurality-Hispanic county AER goes from below 1 to above 1 for the corrected cases across all lookaheads. While the fairness decreases for lookaheads 7 and 14, it increases for 21 and 28; and these results were all statistically significant except for lookahead 21 (see Table \ref{tab:mwu-dynamic} in the Appendix.) 
The AERs for plurality-Asian counties start far below 1 for the baseline and increase to above 1 for all lookaheads. However, there are only 5 plurality-Asian counties in this analysis, so caution is warranted in drawing any conclusions.

For Method 1, we cannot directly compare the baseline and corrected PBL numerical values directly because of the vastly different scale of the predicted values (Method 1 correction produces a case ``trend''). However, we can compare trends within baseline and corrected PBL across groups. Figures \ref{fig:dynamic-PBL-l14-maj} and \ref{fig:dynamic-PBL-l14-plu} show the average PBL distributions for lookahead 14 for majority and plurality, respectively.  
We observe how the relative distributions of PBLs between groups reflect the AERs. For instance, for majority-Black counties baseline PBLs are significantly lower than for majority-White counties, while for corrected cases the distribution of PBLs between the two groups of counties are much closer. This corresponds to the increase in AER from baseline to corrected cases for majority-Black counties \textit{i.e.,} the burden of the prediction error shifts from unprotected to protected groups. Other lookaheads produced similar results (see Figure \ref{fig:pbl-dynamic} in the Appendix).

\begin{figure*}[ht!]
     \hspace{-0.5cm}
     \begin{subfigure}{0.74\textwidth}
        \centering
        \includegraphics[scale=0.5]{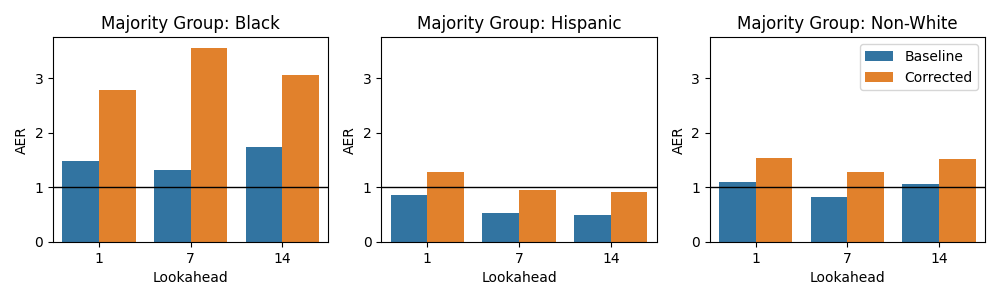}
        \caption{}
        \label{fig:cfr-AER-majority-500}
     \end{subfigure}
     \hspace{-0.5cm}
     \begin{subfigure}{0.24\textwidth}
        \centering
         \includegraphics[scale=0.5]{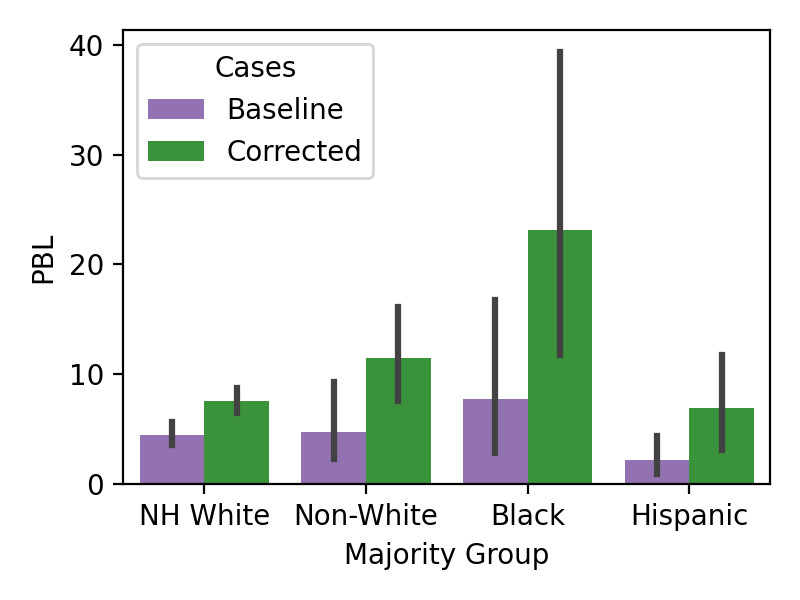}
         \caption{}
         \label{fig:cfr-500-PBL-l14-maj}
     \end{subfigure}
     \caption{Method 2, study period 4/24/20-6/15/20: a) AER values for Majority Black, Hispanic and Non-White counties compared to majority-White counties for both uncorrected and corrected cases. b) The average and 95\% confidence interval of PBL values for each majority group for lookahead 14. PBLs are all multiplied by the median county population (25,726).}
\end{figure*}

\begin{figure*}[ht!]
     \hspace{-0.5cm}
     \begin{subfigure}{0.74\textwidth}
        \centering
        \includegraphics[scale=0.5]{Figures/cfr_AER_Maj_500_cty.png}
        \caption{}
        \label{fig:cfr-AER-plurality-500}
     \end{subfigure}
     \hspace{-0.5cm}
     \begin{subfigure}{0.24\textwidth}
        \centering
         \includegraphics[scale=0.5]{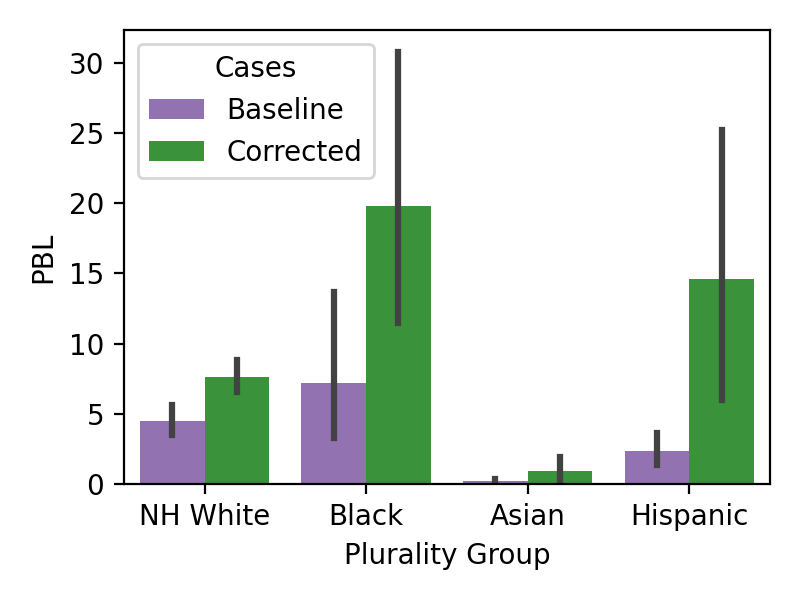}
         \caption{}
         \label{fig:cfr-500-PBL-l14-plu}
     \end{subfigure}
     \caption{Method 2, study period 4/24/20-6/15/20: a) AER values for Plurality Asian, Black, and Hispanic counties compared to plurality-White counties for both uncorrected and corrected cases. b) The average and 95\% confidence interval of PBL values for each majority group for lookahead 14. PBLs are all multiplied by the median county population (25,726).}
\end{figure*}

\subsection{Method 2: Impact of CFR Benchmark on Quantile Regression Fairness} 
As explained in the evaluation protocol,
we evaluate the impact of the CFR benchmark on model fairness for two study periods: from 4/7 to 5/22 and from 4/24 and 6/15. Next we describe the results for each period. 

\subsubsection{Study Period: 4/7/20 to 5/22/20}

For the earlier time period, the resulting AERs for the majority-assignment are shown in Figure \ref{fig:cfr-AER-majority-225}. 
When comparing the baseline against the corrected AER for majority-Black counties, the AER is almost constant for lookahead 1, while increasing from baseline to corrected cases from below one to above one for lookaheads 7 and 14. This result reveals that
applying the case correction method does not change much the fairness, with AER values still close to one. 
However, 
while the baseline models appear to be associated with 
lower prediction errors for majority-Black counties when compared to majority-White (AER$<$1), the correction had the opposite effect, 
although the fairness did not change much. 
These results are validated by our statistical analysis, which shows no statistically significant difference between majority-Black and majority-White counties for the baseline or corrected models (see Table \ref{tab:mwu-cfr-225} in the Appendix).

For Non-White counties, the opposite trend occurs, with baseline AERs above 1 and corrected AERs below one for all three lookaheads
\textit{i.e.,} the predictions with corrected cases output lower prediction errors for majority-Non-White counties when compared to majority-White. On the other hand, while the AER values get closer to one for lookahead 1, they move slightly further away from one for the other two lookaheads, pointing to a decrease in fairness (although not statistically significant). Lastly, for majority-Hispanic counties the AER gets closer to one from baseline to corrected cases for lookaheads 1 and 7, pointing to an improved fairness; and the AER decreases for lookahead 14, which revealed a significant decrease in fairness (see Table \ref{tab:mwu-cfr-225} in the Appendix). All AERs stayed below 1, pointing to lower errors for majority-Hispanic counties when compared to majority White counties.

Figure \ref{fig:cfr-AER-plurality-225} illustrates the baseline and corrected AERs for the plurality assignment. For Black-plurality counties, the AER values are all above one for both baseline and corrected cases.
However, the AER values in the corrected models are larger for lookaheads 7 and 14, pointing to a decrease in fairness, which was also confirmed by our statistical analysis for lookahead 7 (see Table \ref{tab:mwu-cfr-225} in the Appendix).
Across all lookaheads for Hispanic-plurality counties, the AER decreases from above 1 to below 1, with AER values for lookaheads 7 and 14 moving further away from 1, pointing to a decrease in fairness; our tests confirmed 
the statistical significant decrease for lookahead 14. Interestingly, that decrease in fairness points to lower prediction errors for plurality-Hispanic when compared to plurality-White. 
The AER for the 2 Asian counties is below 0.25 in all cases, pointing to smaller errors for plurality-Asian counties when compared to plurality-White both for models trained with and without corrected cases. Although given the small number of counties, this result might not be representative.

Unlike with Method 1, we can directly compare PBLs for the baseline and corrected cases, and uncover potential underlying trends in PBLs that are hidden by the AER metric. For instance, an increase in AER for majority-Black counties could be due to either an increase in PBLs for majority-Black counties or a decrease in PBLs for majority-White counties. However, we instead see that PBLs across just about all lookaheads and group assignments increase from baseline to corrected cases (as seen Figure \ref{fig:pbl-cfr_225} in the Appendix). This is due to the fact that Method 2 corrections almost always increase case numbers compared to the baseline. Figures \ref{fig:cfr-225-PBL-l14-maj} and \ref{fig:cfr-225-PBL-l14-plu} show the two exceptions, where case numbers stay constant for majority-Non-White counties and decrease for plurality-Hispanic counties for lookahead 14. Thus, changes in AER are generally due to relative differences in PBL increases between county groups, rather than to PBL changes in opposing directions \textit{i.e.,} the correction method does not generally shift the error burden from unprotected to protected groups, or vice versa.

\textbf{To sum up, compared to Method 1, the picture is foggier as to the effect of case corrections on prediction fairness, with fairness improving, worsening or not changing depending on race, lookahead and race assignment scheme.} Overall, majority-Black is not affected by the corrected data, while plurality-Hispanic and plurality-Black suffer a significant decrease in fairness for lookaheads 14 and 7, respectively, when corrected cases are used.
Given that study period only considered 226 counties, we posit that the following analysis with 515 counties should provide a clearer picture.

\subsubsection{Study Period: 4/24/20 to 6/15/20}

For the later study period, the results for the majority assignment are shown in Figure \ref{fig:cfr-AER-majority-500}. For every racial group and lookahead, AER increases from the baseline to the corrected cases. This is particularly problematic for majority-Black and majority-non-White counties where baseline AERs are close to or above 1, thereby clearly decreasing fairness. For majority-non-White counties and lookahead 14, this decrease in fairness was confirmed by the statistical analysis, which identified a lack of significant difference in PBLs for the baseline cases and then a significant difference appearing for the corrected cases, as shown in Table \ref{tab:mwu-cfr-500} in the Appendix. For majority-Hispanic counties, the baseline AERs start under 1, and the increase brings AERs closer to 1 for lookaheads 7 and 14 (and not much farther from 1 for lookahead 1), pointing to an improvement in fairness. 

The results are similar for the plurality assignments, shown in Figure \ref{fig:cfr-AER-plurality-500}. For plurality-Black counties, baseline AERs start above 1 and further increase for corrected cases, revealing that predictive models trained with corrected cases can decrease the fairness of the predictions. Statistical tests show that PBLs start off significantly higher for baseline cases for plurality-Black counties and remain significantly higher for corrected cases (see Table \ref{tab:mwu-cfr-500} in the Appendix). A similar trend occurs for plurality-Hispanic counties for lookahead 1.  For plurality-Hispanic lookaheads 7 and 14, AER starts close to 0.5 and moves to the 1.5-2 range. These numbers point to a small decrease in fairness for the corrected model (with AER values slightly farther away from 1 than the baseline), although these differences were not found to be statistically significant. 
AER starts and remains low for the plurality-Asian counties with negligible changes.

Similarly to the earlier time period, PBLs increase across all groups (and as seen in Figure \ref{fig:pbl-cfr_500} in the Appendix, across all lookaheads and group assignment methods). AERs change between baseline and corrected cases because of the different rates of increase in PBL, and not due to shifts in error burden between protected and unprotected groups. For example, the increase in PBL for majority and plurality Hispanic counties is greater compared to majority and plurality White counties, leading to an increased AER between baseline and corrected case loads (see Figures \ref{fig:cfr-500-PBL-l14-maj} and \ref{fig:cfr-500-PBL-l14-plu}).

\textbf{To summarize, compared to Method 1 and the earlier time period for Method 2, the corrections unambiguously decrease the fairness for majority and plurality-Black counties and for majority-Non-White counties. For majority-Hispanic counties, fairness improves, although for plurality-Hispanic, fairness decreases slightly.}

\section{Discussion}

Through different combinations of lookaheads and racial/ethnic group assignments, we see a variety of trends in the impact of case correction methods on downstream model fairness. For Method 1, AERs generally trend towards 1 from baseline to corrections, clearly maintaining or improving the prediction fairness. For Method 2, AERs equally as often trend away from 1 as they do towards 1, with the case correction often times resulting in a decrease in fairness. 
While AER changes between lookaheads within the same correction method and race assignment, there is no consistent correlation between lookahead and AER.  Similarly, using majority versus plurality race assignment has different impacts depending on the method. For example, AERs are similar for majority- versus plurality-Black counties for Method 1, while AERs are quite different between majority- versus plurality-Hispanic counties for both study periods in Method 2.
These differences show the importance of breaking out results by lookahead and race assignment to audit the fairness impact of each method.

It is also worth discussing the normative implications of our fairness metric. We consider a fair model to have similar error rates across different county types. However, given the outsize burden of COVID-19 on marginalized racial and ethnic groups, is the concern for bias as strong when, for example, majority-Black counties have consistently lower error rates than majority-White counties? Regardless, our analysis reveals cases where increases in AERs for non-White majority and plurality counties lead to unfairness, which is definitely a cause for concern.

\textbf{Limitations.} 
There are several limitations and thus room for future work to our paper. Firstly, we examine just two methods for case correction out of many more developed; however, we believe that this analysis is sufficient to show that other methods could suffer from similar biases and this area of research as a whole needs further auditing, prior to incorporation of correction methods into decision-making systems. Secondly, while we attempt to maintain the context of these two methods as best as possible, it is still not an exact replication. Contextual changes include the study time period and regional granularity (county versus country). Thirdly, our analysis occasionally suffers from low sample sizes, particularly for method 2, but also for plurality-Asian counties across both methods. Future work can investigate other group assignment methods, such as assigning labels based on whether each county has above or below the U.S. average for a particular racial/ethnic group \cite{gursoy2022error}. Lastly, we consider just quantile regression methods for COVID-19 prediction for simplicity and interpretability, but other methods, particularly epidemiological SEIR-based methods, may 
show different fairness outcomes. Future work can focus on comparing additional COVID-19 prediction models, including state of the art models that incorporate further data sources, such as human mobility data or masking rates. Regardless, the chosen evaluation framework provides evidence of potential bias.

\section{Conclusion}

Correcting undercounting of COVID-19 caseloads is important for better understanding the spread of COVID-19 and making informed individual or policy decisions regarding risk levels. \textbf{However, our analysis shows that the potential benefits of case correction are often distributed unequally among racial and ethnic groups.} While models built on uncorrected cases may start out biased, it critical for correction methods not to exacerbate this bias.

\bibliography{bibliography}

\appendix

\section{Appendix}

\setcounter{table}{0}
\renewcommand{\thetable}{A\arabic{table}}

\setcounter{figure}{0}
\renewcommand{\thefigure}{A\arabic{figure}}

\begin{table*}[]
\centering
\begin{tabular}{rlllll}
\toprule
Lookahead & Assignment Type & Group & Outcome & Baseline MWU & Corrected MWU \\
\midrule
7 & Majority & Black & more fair & *0.0 & 0.091 \\
7 & Majority & Hispanic & less fair & 0.256 & *0.004 \\
7 & Majority & Non-White & more fair & *0.0 & 0.189 \\
7 & Plurality & Asian & more fair & *0.0 & 0.095 \\
7 & Plurality & Black & both unfair & *0.0 & *0.049 \\
7 & Plurality & Hispanic & less fair & 0.657 & *0.001 \\
14 & Majority & Black & both unfair & *0.0 & *0.029 \\
14 & Majority & Hispanic & less fair & 0.965 & *0.043 \\
14 & Majority & Non-White & more fair & *0.0 & 0.891 \\
14 & Plurality & Asian & more fair & *0.0 & 0.150 \\
14 & Plurality & Black & both unfair & *0.0 & *0.009 \\
14 & Plurality & Hispanic & less fair & 0.629 & *0.014 \\
21 & Majority & Black & both unfair & *0.0 & *0.007 \\
21 & Majority & Hispanic & both fair & 0.389 & 0.234 \\
21 & Majority & Non-White & more fair & *0.0 & 0.075 \\
21 & Plurality & Asian & more fair & *0.0 & 0.327 \\
21 & Plurality & Black & both unfair & *0.0 & *0.001 \\
21 & Plurality & Hispanic & both fair & 0.164 & 0.137 \\
28 & Majority & Black & both unfair & *0.0 & *0.007 \\
28 & Majority & Hispanic & both fair & 0.126 & 0.399 \\
28 & Majority & Non-White & both unfair & *0.0 & *0.012 \\
28 & Plurality & Asian & more fair & *0.0 & 0.335 \\
28 & Plurality & Black & both unfair & *0.0 & *0.001 \\
28 & Plurality & Hispanic & more fair & *0.049 & 0.327 \\
\bottomrule
\end{tabular}
\caption{\textbf{Mann Whitney U tests for Method 1. MWU columns indicate the p-values for the Mann Whitney U test between majority/plurality-White counties and the counties assigned to each row's Group column. The Outcome column indicates whether significance levels changed from baseline cases to corrected cases.}
For each lookahead and each group, we compare the PBLs of the counties of that group's majority (plurality) to the PBLs of the Non-Hispanic White majority (plurality) counties. We consider the two groups statistically significantly different if the Mann Whitney U test comparing these two groups of PBLs is $p<0.05$. The PBLs for the baseline and corrected cases are tested separately. Our framework identifies as ``unfair'' if the two groups are significantly different (and ``fair'' if they are not). Because we are interested in whether corrected cases exacerbate or improve COVID-19 prediction bias, we focus on whether the statistical significance has changed between the baseline and corrected model. Thus, in the ``Outcome'' column for these tables, we highlight whether both baseline and corrected are significant (``both unfair''), both are not significant (``both fair''), or if a change has occurred. If a change has occurred where the baseline is not significantly different and then the corrected model is, we consider the corrected model to be ``less fair''; if the baseline is significantly different and then the corrected model is not, we consider the corrected model to be ``more fair''.
}
\label{tab:mwu-dynamic}
\end{table*}

\begin{table*}[!t]
    \centering
\begin{tabular}{rlllll}
\toprule
Lookahead & Assignment Type & Group & Outcome & Baseline MWU & Corrected MWU \\
\midrule
1 & Majority & Black & both fair & 0.336 & 0.381 \\
1 & Majority & Hispanic & both fair & 0.162 & 0.160 \\
1 & Majority & Non-White & both fair & 0.688 & 0.086 \\
1 & Plurality & Asian & both fair & 0.060 & 0.141 \\
1 & Plurality & Black & both unfair & *0.01 & *0.03 \\
1 & Plurality & Hispanic & both fair & 0.578 & 0.086 \\
7 & Majority & Black & both fair & 0.437 & 0.189 \\
7 & Majority & Hispanic & both fair & 0.090 & 0.137 \\
7 & Majority & Non-White & both fair & 0.726 & 0.061 \\
7 & Plurality & Asian & both unfair & *0.007 & *0.025 \\
7 & Plurality & Black & less fair & 0.071 & *0.033 \\
7 & Plurality & Hispanic & both fair & 0.685 & 0.070 \\
14 & Majority & Black & both fair & 0.818 & 0.419 \\
14 & Majority & Hispanic & less fair & 0.490 & *0.024 \\
14 & Majority & Non-White & both fair & 0.851 & 0.091 \\
14 & Plurality & Asian & less fair & 0.074 & *0.017 \\
14 & Plurality & Black & both fair & 0.107 & 0.072 \\
14 & Plurality & Hispanic & less fair & 0.685 & *0.025 \\
\bottomrule
\end{tabular}
    \caption{\textbf{Mann Whitney U tests for Method 2, earlier study period. MWU columns indicate the p-values for the Mann Whitney U test between majority/plurality-White counties and the counties assigned to each row's Group column. The Outcome column indicates whether significance levels changed from baseline cases to corrected cases.}
For each lookahead and each group, we compare the PBLs of the counties of that group's majority (plurality) to the PBLs of the Non-Hispanic White majority (plurality) counties. We consider the two groups statistically significantly different if the Mann Whitney U test comparing these two groups of PBLs is $p<0.05$. The PBLs for the baseline and corrected cases are tested separately. Our framework identifies as ``unfair'' if the two groups are significantly different (and ``fair'' if they are not). Because we are interested in whether corrected cases exacerbate or improve COVID-19 prediction bias, we focus on whether the statistical significance has changed between the baseline and corrected model. Thus, in the ``Outcome'' column for these tables, we highlight whether both baseline and corrected are significant (``both unfair''), both are not significant (``both fair''), or if a change has occurred. If a change has occurred where the baseline is not significantly different and then the corrected model is, we consider the corrected model to be ``less fair''; if the baseline is significantly different and then the corrected model is not, we consider the corrected model to be ``more fair''.
    }
    \label{tab:mwu-cfr-225}
\end{table*}

\begin{table*}[!t]
    \centering
\begin{tabular}{rlllll}
\toprule
Lookahead & Assignment Type & Group & Outcome & Baseline MWU & Corrected MWU \\
\midrule
1 & Majority & Black & both unfair & *0.002 & *0.0 \\
1 & Majority & Hispanic & both fair & 0.450 & 0.926 \\
1 & Majority & Non-White & both fair & 0.170 & 0.339 \\
1 & Plurality & Asian & both unfair & *0.002 & *0.018 \\
1 & Plurality & Black & both unfair & *0.0 & *0.001 \\
1 & Plurality & Hispanic & both fair & 0.919 & 0.457 \\
7 & Majority & Black & both unfair & *0.001 & *0.0 \\
7 & Majority & Hispanic & both fair & 0.355 & 0.667 \\
7 & Majority & Non-White & both fair & 0.513 & 0.387 \\
7 & Plurality & Asian & both unfair & *0.015 & *0.044 \\
7 & Plurality & Black & both unfair & *0.0 & *0.0 \\
7 & Plurality & Hispanic & both fair & 0.875 & 0.873 \\
14 & Majority & Black & both unfair & *0.013 & *0.0 \\
14 & Majority & Hispanic & both fair & 0.215 & 0.920 \\
14 & Majority & Non-White & less fair & 0.759 & *0.028 \\
14 & Plurality & Asian & more fair & *0.001 & 0.074 \\
14 & Plurality & Black & both unfair & *0.0 & *0.0 \\
14 & Plurality & Hispanic & both fair & 0.580 & 0.318 \\
\bottomrule
\end{tabular}
    \caption{\textbf{Mann Whitney U tests for Method 2, later study period. MWU columns indicate the p-values for the Mann Whitney U test between majority/plurality-White counties and the counties assigned to each row's Group column. The Outcome column indicates whether significance levels changed from baseline cases to corrected cases.}
For each lookahead and each group, we compare the PBLs of the counties of that group's majority (plurality) to the PBLs of the Non-Hispanic White majority (plurality) counties. We consider the two groups statistically significantly different if the Mann Whitney U test comparing these two groups of PBLs is $p<0.05$. The PBLs for the baseline and corrected cases are tested separately. Our framework identifies as ``unfair'' if the two groups are significantly different (and ``fair'' if they are not). Because we are interested in whether corrected cases exacerbate or improve COVID-19 prediction bias, we focus on whether the statistical significance has changed between the baseline and corrected model. Thus, in the ``Outcome'' column for these tables, we highlight whether both baseline and corrected are significant (``both unfair''), both are not significant (``both fair''), or if a change has occurred. If a change has occurred where the baseline is not significantly different and then the corrected model is, we consider the corrected model to be ``less fair''; if the baseline is significantly different and then the corrected model is not, we consider the corrected model to be ``more fair''.
    }
    \label{tab:mwu-cfr-500}
\end{table*}


\begin{figure*}
    \centering
    \includegraphics[scale=0.45]{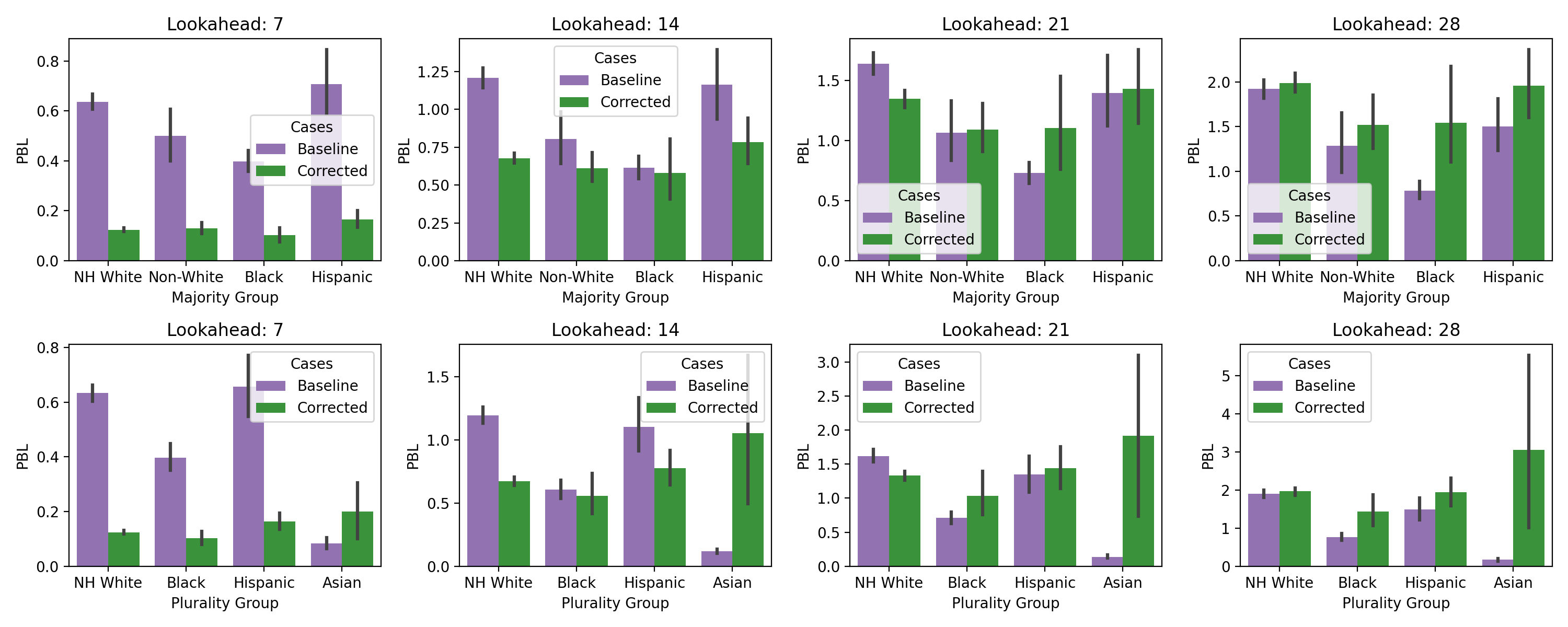}
    \caption{Method 1: The average and 95\% confidence interval of PBL values, for each lookahead and each group assignment. Baseline PBLs are all multiplied by the median county population (25,726) and corrected PBLS are multiplied by 10 for readability.}
    \label{fig:pbl-dynamic}
\end{figure*}

\begin{figure*}
    \centering
    \includegraphics[scale=0.5]{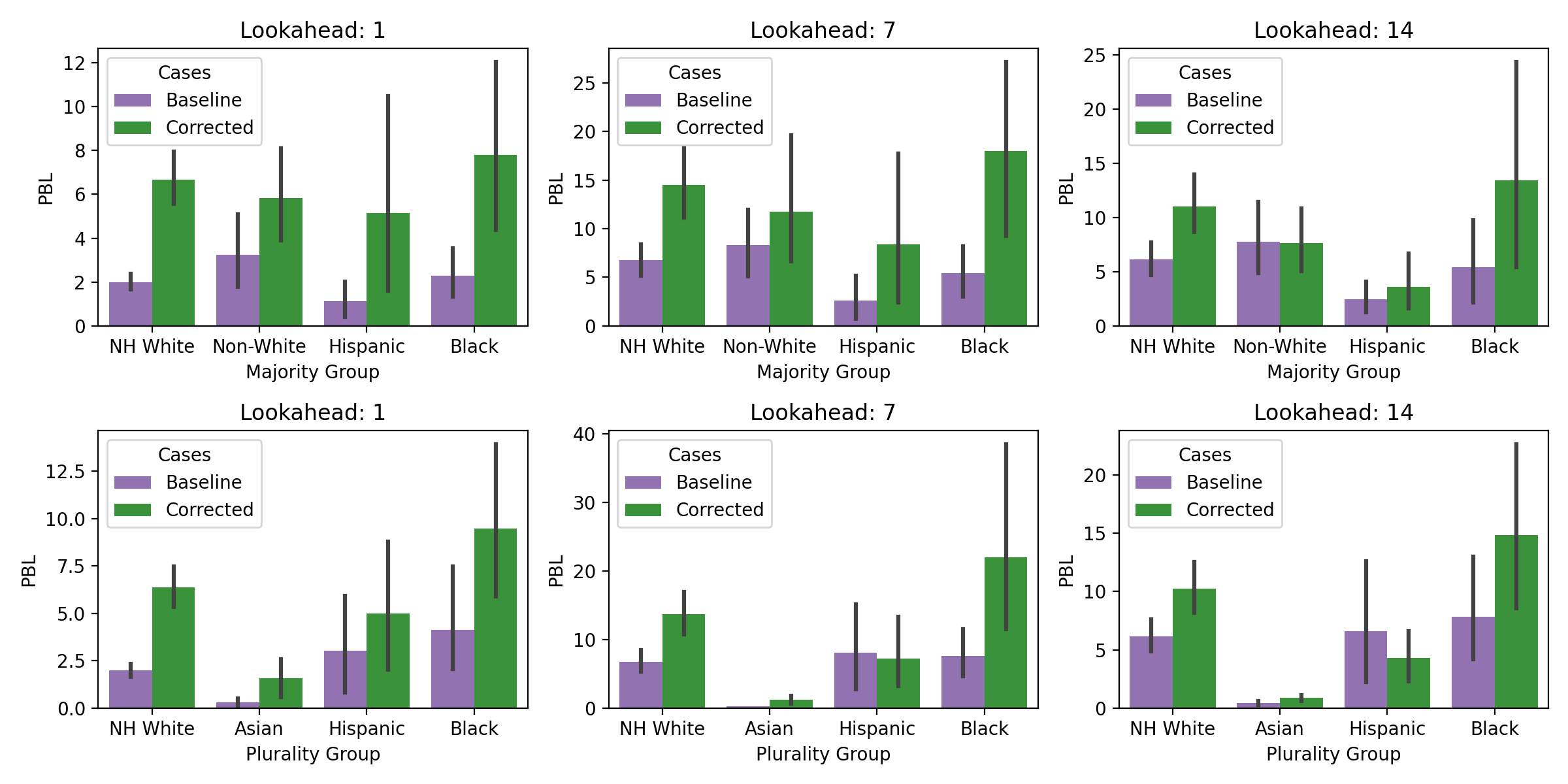}
    \caption{Method 2 earlier study period: The average and 95\% confidence interval of PBL values, for each lookahead and each group assignment. PBLs are all multiplied by the median county population (25,726).}
    \label{fig:pbl-cfr_225}
\end{figure*}

\begin{figure*}
    \centering
    \includegraphics[scale=0.5]{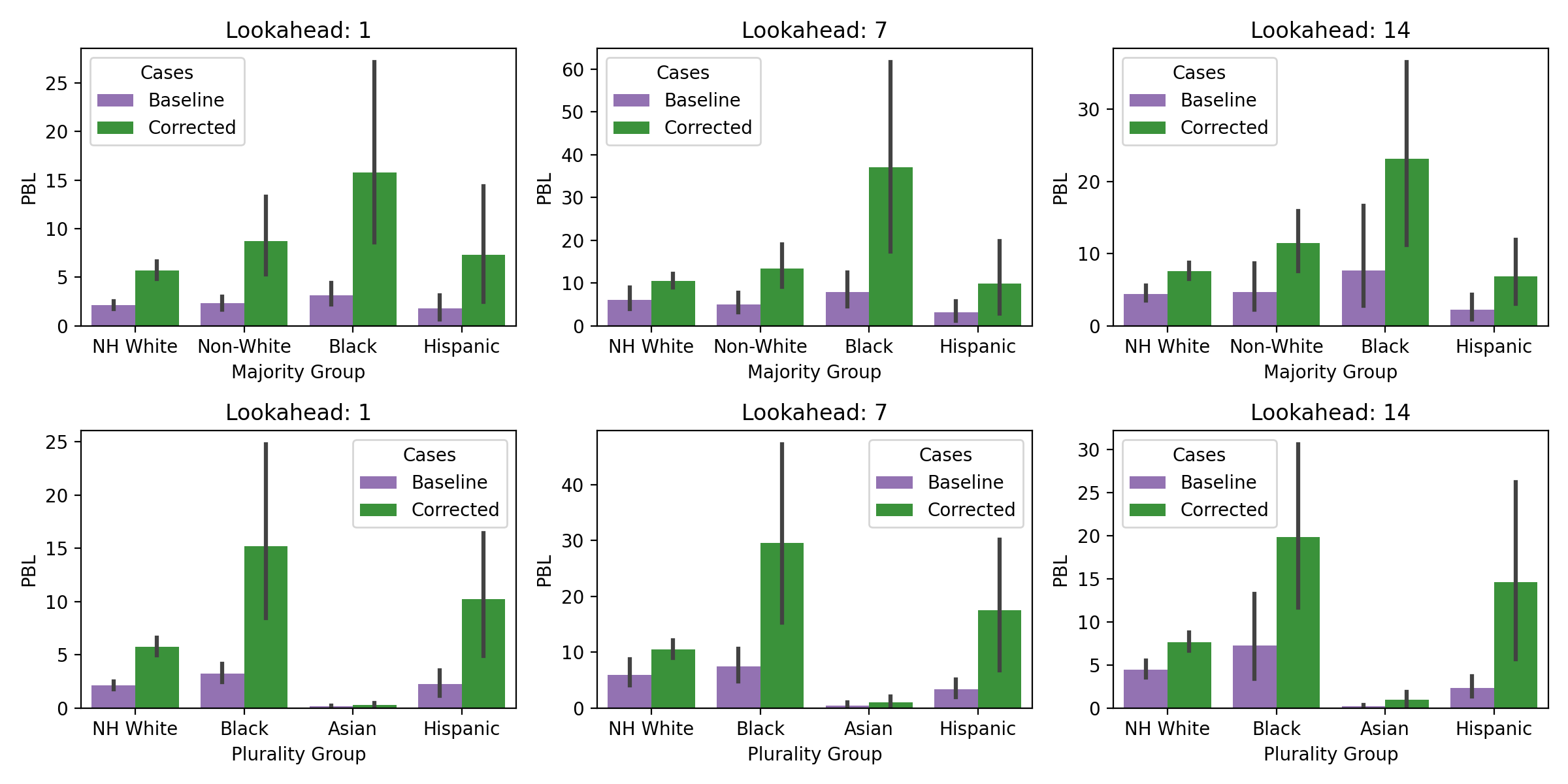}
    \caption{Method 2 later study period: The average and 95\% confidence interval of PBL values for each group, for each lookahead and each group assignment. PBLs are all multiplied by the median county population (25,726).}
    \label{fig:pbl-cfr_500}
\end{figure*}

\end{document}